%% file: ms.tex
\documentclass[
  manuscript= {{Technical Note}}  
]{ailecs-technote}
\pdfoutput=1 
 
\pubdate{March 21, 2022} 
\reportno{TN22/01} 
\copyrightstmt{}

\title{The Data Airlock: infrastructure for restricted data informatics}
\shortheadertitle{The `Data Airlock'.}

\author{Gregory Rolan}
\affiliation{AiLECS Lab, Monash University, Melbourne, Australia}
\email{greg.rolan@monash.edu}
\author{Janis Dalins}
\affiliation{Australian Federal Police, Canberra, Australia}
\author{Campbell Wilson}
\affiliation{AiLECS LAb, Monash University, Melbourne, Australia}

\usepackage{amsmath}

\usepackage{booktabs,microtype,siunitx}
\makeatletter 
\renewcommand{\@seccntformat}[1]{\csname the#1\endcsname\hspace{7pt}}
\makeatother 

\begin{document}

\input{content}


\printbibliography

\end{document}

%% file: content.tex
\begin{abstract}
Data science collaboration is problematic when access to operational data or
models from outside the data-holding organisation is
prohibited, for a variety of legal, security, ethical, or practical reasons. 
There are significant data privacy challenges when performing collaborative data science work 
against such \emph{restricted} data. In this paper we describe a range of causes and risks 
associated with restricted data along with the social, environmental, data, and 
cryptographic measures that may be used to mitigate such issues. We then show how 
these are generally inadequate for restricted data contexts and introduce 
the 'Data Airlock' --- secure infrastructure that facilitates 'eyes-off' 
 data science workloads. After describing our use-case we detail the architecture and implementation of a first, single-organisation
version of the Data Airlock infrastructure. We conclude with outcomes and learning
from this implementation, and outline requirements for a second, federated
version.

\end{abstract}
\section{Introduction}\label{introduction}

The successful application of data science to operational data requires
collaborative effort across the workflow pipeline for the
development, training, testing, debugging, and integration of models and
their outputs \cite{viaene_data_2013}. Increasingly, and particularly
in the public sector, many such collaborators are external to the
data-holding organisation \cite{mikhaylov_artificial_2018}.
Importantly, each of these participants may require access
to the underlying data, the models being developed, and/or any outputs.
Moreover, many grand challenge problems of our time require models to be
trained on data from multiple organisations
\cite{deshpande_revolutionizing_2020} --- in particular, to mitigate
against forms of bias due to narrow datasets
\cite{roselli_managing_2019} .

The challenges of such collaborative R\&D, are vastly more
difficult if there are constraints on such access due to sensitivities
in the data or models themselves. Consequently, a range of techniques
have been adopted for collaborative data science involving sensitive
data or models \cite{jabine_procedures_1993}. Such measures have met
with varying degrees of success depending on the sensitivities involved,
the nature of the workload, and stakeholder needs.

Even so, external collaboration remains problematic when access to
operational data or models is prohibited, for a variety of legal, security,
ethical, or practical reasons. In this paper we describe a new infrastructure for facilitating
`eyes-off' access to restricted data in the pursuit of collaborative
data science research and development. Named `Data Airlock', this
infrastructure was developed in conjunction with a national law
enforcement agency, with a view to facilitate collaboration in the
development of data classification models. 

We begin this paper by exploring various measures for protecting 
sensitive data that remain problematic when applied to restricted data.
We then present the motivating use case for our work --- the development
of automated classifiers of child sexual abuse material (CSAM).
Having provided this background, we introduce the 'Data Airlock', 
infrastructure that addresses the issues with restricted data, and reflect
on our learning from its implementation. We conclude with a 
prognosis for this infrastructure for restricted data informatics.

\section{Background}\label{background}

While domain experts and data scientists play a leading role in
data science collaboration \cite{viaene_data_2013}, a wide variety of
other stakeholders are also involved \cite{commonwealth_of_australia_data_2016, passi_trust_2018, bhatt_machine_2020}.
Increasingly, and particularly within the public sector, organisations
leverage in-house operational data and domain expertise through
partnerships with external researchers, contributors, and service
providers \cite{mikhaylov_artificial_2018}. Similarly, the development
of public sector models in response to societal grand
challenges requires access to a breadth of data that encompasses the 
often granular structure of the public sector and beyond
\cite{mikhaylov_artificial_2018, popa_training_2020, peiffer-smadja_machine_2020}.

Cross-organisational involvement increases the complexity of managing collaboration
\cite{ sun_mapping_2019}, while complicating access and requirements
for security, audit, and transparency \cite{passi_trust_2018}. Moreover 
operational data and requirements change through time, as do the differing objectives, 
responsibilities, and performance expectations of stakeholders, resulting
in the need for continual development, verification, and explanation of models.
These difficulties are exacerbated when operational data and models possess
one or more potentially dynamic sensitivities that place limitations on sharing
of material for collaboration \cite{shokri_privacy-preserving_2015} .

Such sensitivities arise for a number of reasons and may derive from
regulation, policy, or even community expectations. Privacy protections for
subjects recorded in data sets, for example in the health or social
sciences, are often mandated by law \cite{greenleaf_global_2019}.
Such provisions include the protection of personal details as well as information regarding
membership of some data class (or not) based on personal attributes
\cite{li_membership_2013}. Similarly, there may be constraints on
secondary uses of data in the absence of explicit consent 
\cite{custers_click_2016}. Commercial considerations may preclude the open sharing of
proprietary datasets and/or models. In some cases access to data and/or
models may be \emph{restricted} by legislation --- for example in the
law enforcement, national security, and defence sectors. Across this
spectrum of sensitivities, a range of social and technical measures are
available that can ameliorate access constraints.

\subsection{Dealing with sensitive
material}\label{dealing-with-sensitive-material}

The protection of sensitive material can be considered in terms of
security \emph{threats} of two types: \emph{internal} and
\emph{external} \cite{jouini_classification_2014}. An internal threat
in this context is the inadvertent exposure or intentional sharing of
material by a party who has been granted access. Conversely, an
external threat is the deliberate exfiltration, reconstruction, or
dissemination of material by a third-party without permission. The
challenges of sharing sensitive data for analytical work is not a new
problem and predates the data science era. For example, twenty-five years ago, Jabine
comprehensively detailed various causes and characteristics of data 
sensitivities together with measures employed to mitigate sharing issues
\cite{jabine_procedures_1993}. With minor modifications, such
measures have been applied to the collaborative data science problem in
terms of data and/or the models themselves. These measures can be
considered to fall into three broad categories: \emph{social measures},
\emph{environmental constraints}, and \emph{anonymisation methods}. Depending
on the sensitivities concerned, measures from one or more of these
categories may be required to ensure compliance with data assurance
mandates and expectations.

\textbf{Social measures} address internal threats via a range of permissions
and/or structural organisational changes \cite{jabine_procedures_1993}. 
Examples of these include security clearances, non-disclosure agreements,
and procedural or legal dispensations that facilitate or broaden data sharing.
More complex measures include the secondment of individuals between organisations,
or the reassignment of jurisdictional responsibilities to shift data 
custodianship and render sensitivities moot. In some contexts, sensitivities
may be reduced though the explicit solicitation of consent for data sharing
or --- depending on context and prevailing regulation --- a waiver of
some privacy rights at the time of data collection.

\textbf{Environmental constraints} aim to limit the physical contexts of
access to prevent the (internal) inadvertent exposure, intentional
sharing, or (external) theft of sensitive material beyond the limits
imposed by social measures. Material may be encrypted both during
transmission to the access context and at rest when not in active use. 
More significant has been the development of \emph{data safe havens}; 
secure analytics environments that comprise ``appropriate technical and
governance controls which are effectively audited and are viewed as
trustworthy by diverse stakeholders'' \cite[p.~3243]{burton_data_2015}. 
In this case, sensitive material is typically copied into the data safe 
haven (often managed by a trusted third-party institution such as a university)
where suitably authorised stakeholders can remotely access data. Examples 
of such data safe havens include the UK's Secure Anonymised Information Linkage
(SAIL) Databank of health and other public service data \cite{jones_profile_2019}, 
and the Australian E-Research Institutional Cloud Architecture (ERICA)
\cite{australian_research_data_commons_secure_2020}.

Data safe havens employ physical, technological, and procedural,
mechanisms to manage data in a
way that can be ``viewed as safe and trustworthy by all key
stakeholders'' \cite[p.~2345]{burton_data_2015}. Physical safeguards
include strict control of locations where material may be prepared, or
up/down-loaded. Technical safeguards may include dynamically provisioned,
virtualised, and sand-boxed workspaces; multi-factor authentication; and 
detailed activity audit logging. Procedural safeguards may include access
agreements; standardised work flows; vetting of outputs prior to their
release from workspaces, and sanctions for breaches of agreed conduct. 
However, despite these measures, the safe haven approach is predicated on direct access to data, and is difficult to comprehensively protect against the broad range of internal and external threats
\cite{peisert_security_2017, culnane_not_2020}.

\textbf{Anonymisation methods} can be employed to mitigate both internal and
external threats in cases where the identity of data subjects cannot be
disclosed, even though the bulk of the material may be shared
\cite{garfinkel_-identification_2015}. At the most basic level, data
may be filtered with individual identifying fields --- for example, the
18 U.S. HIPPA `Safe Harbour' identification fields
\cite{office_for_civil_rights_methods_2012} --- or metadata removed or
masked in order to anonymise records. In some contexts, datasets can be
permuted and `sliced' so that a complete view of the material is not
disclosed \cite{li_slicing_2010}. With numeric data, other methods can be
employed such as top/bottom-coding (removal of identifiable outliers), 
aggregation of similar records, and averaging of within-group values. 
Perturbation approaches such as \emph{differential privacy} involve
the introduction of random error or `noise' that ``addresses the paradox
of learning nothing about an individual while learning useful information 
about a population'' \cite[p.~5]{dwork_algorithmic_2014}, serving
to improve the anonymity of data subjects
\cite{jabine_procedures_1993, garfinkel_-identification_2015}.

Models may also be vulnerable to privacy-disclosing threats. In some cases
analysis of internal model parameters can be used
to reconstruct the original training data \cite{al-rubaie_privacy-preserving_2019}.
Similarly, repeated and carefully-designed queries that leverage
prediction confidence levels of otherwise opaque models can be used to
reverse-engineer training data \cite{fredrikson_model_2015}. Depending
on the nature of data and algorithms concerned, some models themselves
internally implement transforms such as differential privacy to mitigate
such attacks \cite{abadi_deep_2016}. The training task can also be
partitioned into `teacher' and `student' models for knowledge transfer
in a kind of generative adversarial network
\cite{papernot_semi-supervised_2017} that mitigates against
reconstruction and model inversion attacks. However, none of these
measures address the primary restricted data problem.

\subsection{Limitations of sensitive data
measures}\label{limitations-of-sensitive-data-measures}

While such measures can go a long way to protect the sensitivities of
material, the efficacy of their use is often highly contextual and
subject to a range of limitations. Social and environmental measures are
not absolute and may conceivably be circumvented either inadvertently or
deliberately by end-users, or by third-parties with malicious intent
\cite{culnane_not_2020}. For example, reliance on one-time agreements
and vetting may give rise to a false sense of security as, over time,
workflow pipelines may increase in scope and reach; the motivations of
individuals or project partners may change; or end-user equipment may
become compromised by malicious third parties. Similarly, operational data
often undergoes continuous churn. This continual data 'drift' means that
models need to be continually updated and, possibly
re-worked; further straining `one-shot' paradigms of sensitivity
analysis. Technological measures may fail due to misunderstanding or
misapplication \cite{culnane_vulnerabilities_2017}, or through poor
framing of risks (e.g., where  output controls are circumvented by 
'screen-capture' mechanisms). Continuous risk-based assessment
may be more effective than static designs based on measures of `safety'
\cite{culnane_not_2020}.

The anonymisation methods described above are a trade-off between the
anonymity of data subjects and the fidelity of the data
\cite{garfinkel_-identification_2015}. In some contexts, obscured data elements may compromise data science workloads. Moreover,
such transforms, may not be applicable or practical for other types of data such as text,
image, audio, or video.  In practice, re-identification of `de-identified' data
may occur via a number of mechanisms either deliberately or
inadvertently by end users, or due to increased scrutiny following a
data breach \cite{el_emam_systematic_2011}.
Together, these factors mean that reliance on anonymisation methods to protect
sensitive data may be misplaced.

Finally, all of these measures are predicated on the assumption that
at least some of the data may be shared in the first place. Whether due to the
above-mentioned concerns, or a-priori restrictions on the disclosure of
data outside an organisational context, there
are many circumstances in which these measures are not applicable for
highly sensitive or restricted data. In cases where direct access to
such data is impractical, unethical, or illegal, \emph{eyes-off} mechanisms
(i.e.~without direct access to data) are needed for the development, testing,
comparison, and integration of analytics models.

\subsection{Dealing with restricted data and
models}\label{dealing-with-restricted-data-and-models}

While many of these social measures and anonymisation methods are insufficient
or inappropriate for use in eyes-off environments, there remain a
number of environmental constraints that may be used to protect against
specific threats.

A form of data safe-haven can be employed that incorporates a
\emph{Trusted Execution Environment} (TEE). TEEs have varying
proprietary hardware implementations, but generally comprise 
cryptographically secure memory and computational partitions that
enable isolation of specific workloads from others running 
on the system, and remote attestation of workload integrity  
\cite{sabt_trusted_2015}. They can be particularly applicable to virtualised
environments, where sensitive workloads may need to be protected from
threats originating from hypervisor hosts and/or other virtual guests.
TEE workloads are often required to be built with environment-specific
software development tool-chains;  initialising and loading workloads 
into isolation partitions as well as unloading outputs upon completion.

However, TEE design has no inherent separation between a trusted
workload and its data. Additional protocols (such as code inspection,
encryption of data, secure encryption key loading, and
so on) are needed to ensure eyes-off access to data by a given
workload \cite{ohrimenko_oblivious_2016}. Additionally, a TEE often
imposes a performance overhead due to a number of factors. In the case
of data science workloads, this overhead can be significant, and can result
in a performance degradation of an order of
magnitude or more \cite{akram_performance_2020}. The trade-offs between
TEE security and these drawbacks along with the difficulties in TEE
implementation \cite{costan_intel_2016} and a
range of (admittedly complex) hardware attack vulnerabilities
\cite{francisco_intels_2020}, mean that a TEE may not yet be an
effective element of an eyes-off data science solution.

Another hardware device that may be employed to mitigate some internal
and external threats is the \emph{Data Diode}. Data diodes are hardware
devices that physically enforce a one-way flow of data between nodes or
networks, ensuring ``that no data can be passed, either explicitly or
covertly, in the opposite direction''
\cite[p.~1]{stevens_implementation_1999}. In this way data may be transmitted 
while assuring both the integrity of the sender and the
impossibility of other data being exfiltrated from the sender via that
channel. While a data diode does not directly support eyes-off
access of data, and has the drawback that human intervention is required
to verify receipts and request re-transmission if required, it may be a
useful component for use in highly secure networks.

A fourth set of measures involves the use of provably \emph{secure
protocols} that facilitate the cooperation of two or more parties in the computation of
results from data held privately by each party \cite{lindell_secure_2009}. While such techniques can address some of the issues surrounding cooperative training or inference, they do not
address the fundamental issues of eyes off access to another party's
data. More promising is \emph{Homomorphic Encryption} \cite{acar_survey_2018}
that enables the computation of functions against
encrypted data that return encrypted results to the querier. These functions offer secure equivalents to functions operating on unencrypted data. If used
with asymmetric key schemes, such techniques can facilitate secure,
eyes-off access --- albeit with limited function primitives and a severe
performance penalty. In relation to
machine learning workloads, this effort has largely focused on inference
tasks against encrypted data (rather than the initial training), and
even then, with simplified data and/or models
\cite{chou_faster_2018,  wang_homo-elm_2020}.

Thus the difficulties of dealing with restricted data, outside of its
operational or custodial context \cite{data_republic_why_2019}, are
compounded by the impracticality of `simple' anonymisation methods that reduce
the utility of data on the one hand, and the inapplicability of complex
transformations to generalised data science workloads the other. What is
required is a third way; infrastructure that facilitates eyes-off
application of data science workloads involving restricted data and/or
models. However, before diving into the architecture of such an
infrastructure, we shall describe our motivating use-case for this work.

\subsection{The CSAM use case}\label{the-csam-use-case}

The motivation for our work is a collaboration with a national law
enforcement agency to develop tools to aid the investigation of Child
Sexual Abuse Material (CSAM). Recent years have seen rampant growth of
production and dissemination of such material
\cite{bursztein_rethinking_2019, jay_investigation_2020}. 
The sheer number of items that need to be analysed, is straining limited law 
enforcement and juridical resources \cite{dalins_laying_2018}.
Additionally, repeated and ongoing exposure to often highly disturbing
imagery by law enforcement, forensic, and juridical officers (and others such as
those performing ICT services or undertaking R\&D in the area) is an increasing
source of secondary trauma \cite{seigfried-spellar_assessing_2018}.
While not entirely removing the need for human analysis, automated
triaging of CSAM goes some way to reducing damaging exposure.

\begin{figure}
\centering
\includegraphics[width=\textwidth]{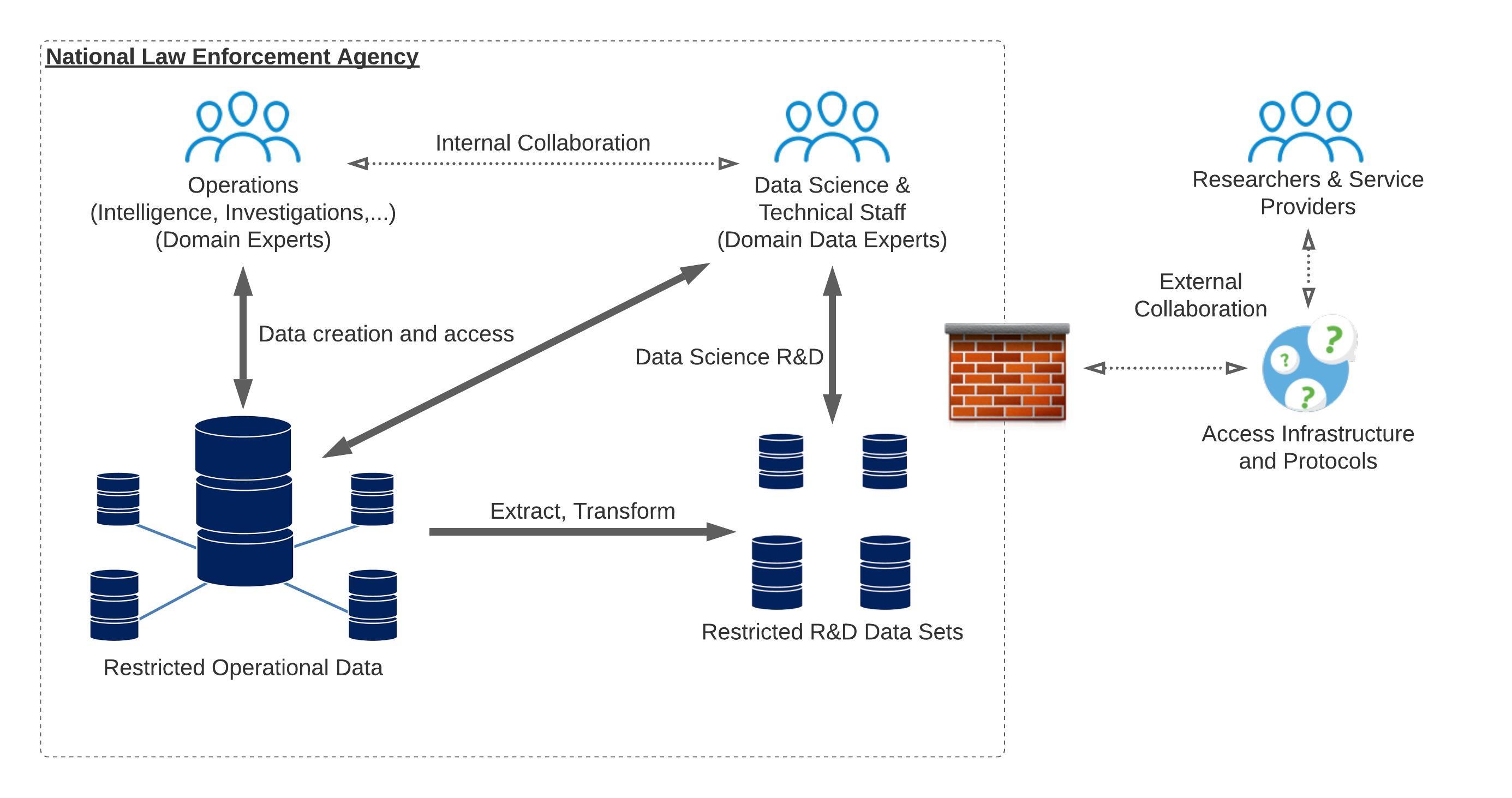}
\caption{Generalised Restricted Data cse case.\label{fig:fig1}}
\end{figure}

Our aim is to develop machine learning models for the automated detection,
identification, and categorising of such material. This is a challenging task,
not only in terms of the technical difficulties
in model development, and the risks associated with working with disturbing
material, but also because of restricted access to this data. Our use case is therefore a
good example of highly sensitive operational data required for ongoing research and development as shown in Figure
\ref{fig:fig1}.

Within the national law enforcement agency, internal 
stakeholders generate and reference operational data as part of their
day-to-day workflow. This sensitive, internal data can take a multitude
of forms including structured data, images, text, video, audio, and binary 
(e.g. device images) along with contextual metadata. These internal 
\emph{domain expert} stakeholders rely on technical staff to develop and 
enhance intelligence and investigative capabilities. These technical 
stakeholders are \emph{domain data experts}, with an
understanding of the formats, volumes, and content of this data together
with an appreciation of the technical and social challenges of working
with it.

The problem arises when those involved in capability development wish to 
collaborate outside the organisation. In our case, the data sets
comprising CSAM are subject to the legislative restriction of any 
``material that depicts or describes activity relating to child sexual
abuse'' \cite{commonwealth_of_australia_criminal_2020}. This led
directly to the creation of infrastructure and protocols that facilitate
such collaboration without the possibility of breaching the legal
restrictions placed on the data.

\section{Initial Data Airlock Design
1.0}\label{initial-data-airlock-design}

This infrastructure was dubbed `Data Airlock' to emphasise the
separation of workloads from restricted data --- in our case,
the CSAM data sets used to train, validate, and test models as they
underwent development. As successful models would
be deployed into law enforcement production, there
was no requirement for this infrastructure to handle the inference case (although model validation was required in addition to training).
Similarly, there was no requirement to deal with model sensitivity.

\subsection{Architecture}\label{architecture}

The high-level architecture for this iteration, shown in Figure
\ref{fig:fig2}, comprises software and hardware components located in
a secure data centre. This infrastructure shares
some similarities with a data safe haven, inasmuch as it is a separate and
partitioned computing platform accessed remotely by users and administrators.
However, the secure connectivity and interaction between the isolating
partitions, as well as the treatment of the restricted data are markedly
different.

The architecture is divided into three logical zones --- the
\emph{public} zone accessed by external R\&D collaborators and
workflow administrators; the high-performance \emph{secure} zone in
which validated workloads run against restricted data; and the
\emph{restricted} zone where encrypted sensitive data is stored. The
three zones operate under different security and access models. In our
implementation, the public zone nodes are virtualised on a single
server. The secure zone is necessarily located on its own high
performance server. The restricted zone runs on a general purpose
computer with high-performance, encrypted storage. 

Public zone access is provided via a web application. Internet access
(web and secure shell) access is only available to the public zone, 
while ssh access to the secure zone (e.g. for system administration) 
is permitted on a separate connection from the public zone. Within the
restricted zone, all manual activity (e.g. software maintenance;
data loading; etc.) takes place via physical, on-site access. The roles of these zones and their connectivity are explained in more
detail as follows:

\begin{figure}
\centering
\includegraphics [width=\textwidth]{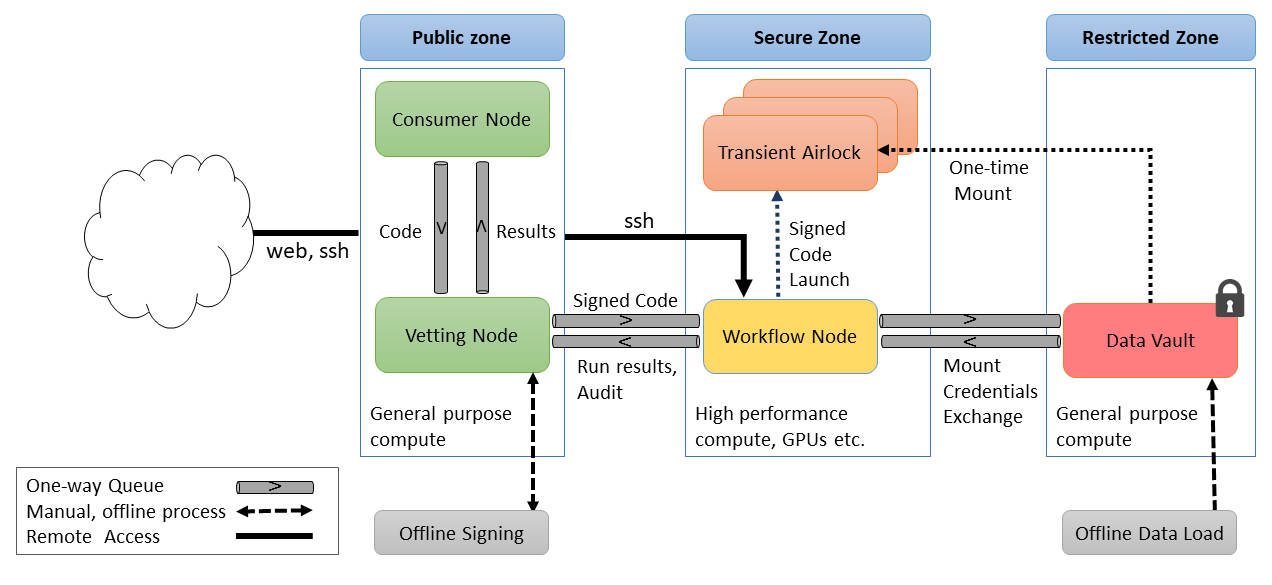}
\caption{Airlock 1.0 Architecture.\label{fig:fig2}}
\end{figure}

\subsubsection{Public Zone}\label{public-zone}

The public zone provides a \emph{Consumer Node} web interface to
external collaborators --- in our case, researchers developing
machine-learning models. This web app is
accessible remotely, and secured using Auth0
\cite{auth0_inc_auth0_2021}. The two consumer functions
are (i) specification of the workload and uploading of code implementing
the model (including pre/post processing stages such as data wrangling; training-validation splits; hyperparameter settings; and so on) to be vetted and subsequently run; and (ii) retrieval of vetted 
results from the run.

As will be described below, the secure zone is implemented on NVIDIA
DGX-1 hardware, enabling the use of pre-built NVIDIA CUDA docker
containers \cite{nvidia_corporation_catalog_2021}. The use of
standardised and externally maintained containers removes the necessity 
for collaborators to replicate the secure zone runtime environment
while  reducing the effort of maintaining runtime environments, and the 
likelihood of environmental security holes. It also enables input
vetting of comparatively small pieces of code, as opposed to whole
containerised runtime environments.

However, this standardisation necessarily introduces constraints for the
model development, limiting the model code to be written in Python (or
code with a Python wrapper and bindings), and forcing the use of
standard Python data science support libraries for model development.
These constraints provided no difficulties for our project, which
employed standard Python model development.

The public zone also provides a \emph{Vetting Node} interface for
 data custodian workflow administrators
--- in our case, data sciences staff from the national police agency. As
with the consumer node, this interface is served by the web app and
employs one-way, persistent queues to interact with other components of
the infrastructure. 

The code vetting is a social measure whereby the job (model and pre/post processing code)
enqueued from the public zone is manually analysed by a workflow
administrator for elements that would, intentionally or inadvertently,
exfiltrate restricted data from the sensitive zone. Once vetted,
a cyrptographic hash of the code is signed by a vetter (using a nonce to
prevent relay attacks) to ensure integrity of code executed in the secure
zone. Importantly, as no private keys are held on the infrastructure, 
the actual signing takes place offline on a vetter private system. Once
signed  the code is then enqueued for execution in the secure zone.

Similarly, results from a job run within the secure zone are enqueued
for review in the vetting node, in this case, manual examination for
evidence of exfiltrated data in the output. If acceptable, the
results are then enqueued for reception by the submitter. These results
retrieved at the consumer node include notification of the failure of
the code to pass vetting; statuses of the job in the various queues;
and/or the outputs of the job in term of reports, data files, or images
(e.g.~performance graphs).

This inherently a batch-oriented system. From the point of consumer submission onward, there is no run-time interaction with the job save for the input/output vetting.

\subsubsection{Secure Zone}\label{secure-zone}

The secure zone contains a high-performance NVIDIA DGX-1 computing node
that runs models on the data in isolated virtual environments or
\emph{airlocks} implemented using Docker
\cite{docker_inc_empowering_2021} containers.

Jobs are dequeued within the sensitive zone by an automated process in
which the code signature is checked using public vetter keys. If
successful, one-time credentials for data access are created and the
appropriate standard NVIDIA docker container is selected. This transient
airlock is then launched with the vetted code as its entry point and the
restricted data is mounted using the one-time credentials. Each airlock
is also provided with ephemeral workspace and output mounts to which it
can write data. At the completion of the job, the restricted data is
unmounted; the resulting model, any processing results, and output logs
are enqueued for vetting; and the temporary mounts are destroyed.

\subsubsection{Restricted Zone}\label{restricted-zone}

Up to this point the Data Airlock infrastructure resembles many aspects of
a typical safe data haven. However, unlike safe-haven sand-boxed environments such as ERICA or SAIL, it is the restricted zone
design that departs from the safe haven concept (albeit with extra layers of potentially manual vetting).

The restricted zone \emph{data vault} provides secure
storage for sensitive data that is physically
loaded on-site by data custodians into volumes encrypted with a manual
boot-time password. The data vault dequeues secure zone requests for
access to restricted data and returns one-time credentials for that
access. The restricted data volume is then made available to the secure
zone when requested, secured by those one-time credentials.

\subsection{Outcomes}\label{outcomes}

The Data Airlock architecture proved successful for our project,
enabling the collaborative development and improvement of CSAM
classifier models without requiring external research staff to directly
assess restricted CSAM material. The important term here is
\emph{collaboration}, as the very nature of restricted data does not
permit the wholesale outsourcing of data science R\&D against `raw'
operational data. Instead, domain data experts within the data custodian
organisation are needed to perform any initial data preparation to
render the raw operational data into a form amenable for model
development. While agreement on data formats, label schemas, and label
element schemes can be reached between all collaborators, this initial
bootstrapping of the data needs to be performed `in-house'. Similarly,
any labelling obviously needs to be performed by the data custodian ---
in our case, labels were derived from criminal investigation work that
had previously assessed and categorised CSAM material.

In our project, some of the data pipeline work was performed as part of
the external workflow in the secure zone --- for example, de-duplication
of images using (restricted) perceptual hash sets \cite{du_perceptual_2020},
downscaling of images, etc. Ultimately,
though, such wrangling needs to be conducted in collaboration with domain data experts
to ensure data quality is maintained. Without the ability to scrutinise
data (and labels), and processing results, the detection of data, labelling, and/or pre-processing
anomalies is difficult. To this end, an improvement to the Data Airlock
workflow would be a formal mechanism for querying or reporting
data/labelling anomalies, changes to data characteristics, and so on.
This requirement points to the necessity for early and ongoing
collaboration with domain data experts to maintain the integrity of
model R\&D as data drifts, requirements change, and domain understanding
improves.

Beyond the data bootstrapping and wrangling issues, this initial
iteration of the architecture also exposed a number of other issues and
opportunities for further improvement.

\subsubsection{Scope and scale}\label{scope-and-scale}

Firstly, as mentioned above, the Data Airlock was only designed to
protect restricted data --- not models. In our case the models being
developed were for use exclusively within the national law enforcement
agency and our researchers were trusted to not disclose models to any
third party. However, in the general case, these social measures may not
be sufficient with the Data Airlock infrastructure needing to protect 
models in addition to data. Scenarios for such use may include the
assured benchmarking or comparison of proprietary models against a
standard (perhaps restricted) dataset; research in hardening models against
threats; and just generally
restricting the possibility of analytical models ``escaping into the
wild'' which could be problematic in law enforcement, security, or
defence contexts.

Similarly, this implementation was not designed to support a plurality of collaborators and
data custodians from multiple organisations. Although this was not an
issue for us, it is certainly a requirement for a more generalised
solution to the restricted data problem. A federated platform for
controlled and configurable `eyes-off' access to restricted data would
enable deep and broad collaboration between a range of disparate
data-holders and researchers for the training, testing, and comparison
of models against data that is held elsewhere.

Such an infrastructure would need to incorporate federated
authentication, authorisation, workflow management, and audit. 
Data-custodians would create catalogue entries of
pre-vetted jobs, tasks, and data recipes etc. that run against their
secure zone compute resources and restricted zone data, and then grant
access to these in much the same manner that vetting is currently
performed. From a collaborator perspective, these pre-vetted jobs,
tasks, and data recipes would be combined to run across multiple
organisation datasets/models in a federated, standardised, and secure
manner.

Another limitation was the web interfaces of the public zone and the
`bare metal' interface of the restricted zone. These gave rise to a lack
of integration of the airlock with data custodian and collaborator
workflows. A more generalised Data Airlock should provide some support for
integration with operational data workflow (perhaps using data diodes);
for example, versioning of datasets, coordination of data maintenance with 
the availability of data for mounting, and so on. Similarly, integration 
with collaborator workflows could be effected through the use of an SDK or API to reduce
(but, importantly, not eliminate) instances of manual intervention by
researchers to submit workloads and receive results.

\subsubsection{Performance improvements}\label{performance-improvements}

Our project exposed a number of other
bottlenecks that could be improved in future versions. For example, the
PKI environment used for code signing etc. required the manual uploading
of encryption keys into the restricted zone system. A superior mechanism
for this could be the use of some form of tamper-resistant hardware for key
management (such as a hardware security module) that would reduce the complexity and improve security of such
maintenance.

Similarly, the vetting of submitted workloads and returned outputs
was an entirely manual activity. It may be that
portions of such vetting could be performed in an automated fashion,
perhaps differentiating boilerplate code and results from bespoke
sections, using machine learning techniques as part of the workflow
infrastructure. This is a research area in its own right, perhaps along the lines
of machine learning anomaly or intrusion detection in high performance
environments described by Peisert \cite{peisert_security_2017}.

Finally, it should be noted that the workload scheduler in this initial
architecture was single-threaded, enabling one isolated airlock at a
time to execute, providing access to the full complement of compute
resources (e.g.~CPU/GPU/RAM) available on the secure zone server. A
generalised architecture should allow for more granular and parallel
execution of isolated airlocks.

\subsubsection{Data mounting
improvements}\label{data-mounting-improvements}

If such parallelism were implemented, then the isolation of
container-based mounts would become a priority for a generalised
solution. Currently, due to standard docker sandboxing, the restricted data mounts are not
privileged and are available to the whole (albeit, currently single
airlock instance-at-a-time) secure zone server. For parallel container
execution, the ability to isolate ephemeral data mounts to particular
transient containers will be necessary.

Along these lines, increasing the granularity of data that is mounted
would both tighten security, and help with the coordination of data
maintenance and the mounting of data for workflows.

In a similar vein to this data granularity, is it possible, though
unlikely, for an intruder in the secure zone to theoretically spoof a
mount request, obtaining access to the content of the data vault. One
protection against this scenario would be to increase the granularity of
signed workflow elements along with code --- in this case, some form of
signed mount request --- that would be checked within the restricted
zone. Of course, this would increase the complexity of vetting and
signing operations, but a semi-automated custodian workflow described
above may offset this impost.

\section{Conclusion}\label{conclusion}

We have discussed how a range of social, environmental, data transform,
and secure protocol measures can address various internal and external threats
to sensitive data, but remain insufficient to protect restricted data in 
collaborative contexts. On the other hand, a new, purpose-built infrastructure, 
architecture dubbed 'Data Airlock' has demonstrated its utility in our project
concerning restricted law enforcement data.

Our experience in deploying and using the Data Airlock has exposed a number of
assumptions and shortcomings in our implementation, leading to a an additional
set of requirements for interoperability and scalability in order to support a
distributed and heterogeneous research community. The design and implementation
of a second version of the Data Airlock is currently underway that will open up 
possibilities across research domains; running or comparing models without
disclosing their technical detail and applying different restricted data 
'recipes' based on access and trust criteria.

Such a highly-secure, federated platform will enable controlled 'eyes-off' access to large, sensitive data sets in order to facilitate collaboration between disparate data-holders and researchers across a variety of problem domains, including law-enforcement, defence, security, medicine, and social services.